\documentclass{aa}

\usepackage{graphics}

\def\t{\thinspace}
\def\la{\mathrel{\hbox{\rlap{\hbox{\lower4pt\hbox{$\sim$}}}\hbox{$<$}}}}
\def\ga{\mathrel{\hbox{\rlap{\hbox{\lower4pt\hbox{$\sim$}}}\hbox{$>$}}}}

\def \sss {\scriptscriptstyle}

\def \epm {e^\pm}
\def \bpc {B_{\rm pc}}
\def \rth {R_{\rm th}}
\def \rpc {r_{\rm pc}}
\def \rns {R_{\rm ns}}
\def \heq {h_{\rm eq}}
\def \hacc {h_{\rm acc}}
\def \hmc {h_{\sss \rm MC}}
\def \scloss {|\dot\gamma_{\sss \rm ICS}|}
\def \resloss {|\dot\gamma_{\rm res}|}
\def \knloss {|\dot\gamma_{\sss \rm KN}|}
\def \gain {\dot\gamma_{\rm acc}}
\def \geql {\gamma_{\rm eq1}}
\def \geqh {\gamma_{\rm eq2}}
\def \mfp  {\lambda_{\sss \rm ICS}}

\def \eps {{\cal \scriptstyle E}}
\def \epr {\epsilon^\prime}
\def \eb  {\epsilon_{\scriptscriptstyle B}}
\def \bcr {B_{\rm cr}}
\def \st  {\sigma_{\scriptscriptstyle T}}
\def \skn  {\sigma_{\scriptscriptstyle KN}}

\begin{document}

\thesaurus{06(02.01.1; 02.18.5; 02.19.2; 08.14.1; 08.16.6)}

\title{ Electron energy losses near pulsar polar caps: \\
a Monte Carlo approach}

\author{J. Dyks \and B. Rudak}
\institute{Nicolaus Copernicus Astronomical Center, Rabia\'nska 8, 
87-100 Toru\'n, Poland\\
E-mail: jinx@ncac.torun.pl, bronek@camk.edu.pl}

\date{Received / accepted }
\titlerunning{Electron energy losses near polar caps}
\authorrunning{Dyks \& Rudak}
\maketitle

\begin{abstract}
We use Monte Carlo approach to study the energetics of electrons 
accelerated in a pulsar polar gap.
As energy-loss mechanisms we consider magnetic Compton scattering of 
thermal X-ray photons and curvature radiation.
The results are compared with previous calculations 
which assumed that changes of electron energy occurred smoothly
according to approximations for the average energy 
loss rate due to the Compton
scattering.

We confirm a general dependence of efficiency of
electron energy losses due to inverse Compton mechanism 
on the temperature and size of a thermal polar cap and on the pulsar
magnetic field. However, 
we show that trajectories of electrons in energy-altitude space 
as calculated in the smooth way do not always coincide with
averaged Monte Carlo behaviour. In particular, for pulsars with 
high magnetic field strength ($\bpc \ga 3\times 10^{12}$ G)
and low thermal polar cap temperatures ($T \la 5\times 10^6$ K)
final electron Lorentz factors computed with the two methods
may differ by a few orders of magnitude. We discuss consequences for 
particular objects with
identified thermal X-ray spectral components like \object{Geminga},
\object{Vela}, and \object{PSR B1055$-$52}.

\keywords{acceleration of particles --
radiation mechanisms: non-thermal -- scattering -- stars: neutron 
-- pulsars: general}

\end{abstract}

\section{Introduction}
Curvature radiation (CR) and magnetic inverse Compton scattering (ICS) 
are usually considered to be the most natural ways of hard gamma-rays
production operating at the expense of pulsar rotational energy
(eg.~Zhang \& Harding \cite{zh}, and references therein).
These two radiation mechanisms dominate within two different ranges of Lorentz
factors $\gamma$ of beam particles (ie.\t those leaving
the polar cap).
When $\gamma \la 10^6$,
magnetic inverse Compton scattering  plays a dominant role in braking beam 
particles (Xia et al.~\cite{xqwh}; Chang \cite{chang}; Sturner \cite{sturner}) 
and is the main source of hard
gamma-ray photons (Sturner \& Dermer \cite{sd}; Sturner et al.~\cite{sdm}).

When $\gamma\ga 10^6$, the curvature radiation becomes responsible for cooling  
beam particles (eg.\t Daug\-her\-ty \& Harding 1982), yet,
the inverse Compton scattering may still be important for secondary $\epm$-pair 
plasma. Zhang \& Harding (\cite{zh}) show that different dependencies of X-ray 
and gamma-ray luminosities on the pulsar spin-down luminosity as 
inferred from observations
by CGRO, ROSAT and ASCA (Arons \cite{arons}; Becker \& Tr\"umper \cite{bt}; 
Saito et al.\t \cite{skks}, respectively) can be well
reproduced when this effect is included.

So far, the influence of ICS on electron energy have been investigated
most thoroughly by Chang (\cite{chang}), Sturner (\cite{sturner}),
Harding \& Muslimov (\cite{hm}), 
and Supper \& Tr\"umper (\cite{supper}). One of several interesting results 
they found was that energy losses due to resonant ICS
can limit the Lorentz factors $\gamma$ of electrons to a value 
which depends on both
electric field strength $E_\parallel$, temperature $T$ and radius $\rth$ of 
thermal polar cap, and on the magnetic field strength $\bpc$ at a polar cap.
For example Sturner (\cite{sturner}) shows that assuming 
the acceleration model of Michel (\cite{michel}) 
and a thermal polar cap size comparable to that determined by the open
field lines, the Lorentz factors
$\gamma$ are limited to $\sim 10^3$
if $\bpc>10^{13}$ G, and $T > 3 \times 10^6$ K.
This acceleration stopping effect becomes more efficient for stronger
magnetic fields, 
and it led Sturner (\cite{sturner}) to propose it as a possible
explanation for an apparent cutoff around 10 MeV in gamma radiation 
of \object{B1509$-$58} (Kuiper et al.~\cite{kuiper}).

Those results were obtained with a numerical method 
which 
assumes smooth changes in electron energy and makes use of 
expressions for \emph{averaged} energy losses due to the magnetic ICS
(Dermer \cite{dermer}).
The continuous changes of the Lorentz factor with height as determined with
this fully deterministic treatment are 
considered as representative for a
behaviour of a large number of electrons. Actually however,
electrons lose their energy in discontinuous scattering events.
Between some of these events (which \emph{occasionally} can be very distant even for 
a short mean free path) the energy of
an accelerated electron may increase considerably (even a few times) 
and therefore, the electron can find itself in quite different conditions
than if it were losing its energy continuously (the mean free path depends
very strongly on $\gamma$). Since the energy loss rate due to the resonant
ICS is not a monotonic function of $\gamma$ this effect may be essential for 
future behaviour of the particle. 

The aim of our paper is to investigate the energetics of electrons
on individual basis with a Monte Carlo treatment. In particular, a
question of the ICS-limited acceleration is adressed.
In Section 2 we calculate electron energy losses near a neutron star
using both, the Monte Carlo approach and then the 
smooth method
(after Dermer 1990; Chang 1995; Sturner 1995) for comparison.
We consider two distinct cases of electron acceleration:
1) in the constant electric field,
proposed by Michel (\cite{michel}) (hereafter M74);
2) in the electric field elaborated by Harding \& Muslimov (\cite{hm})
(hereafter HM98).  
Section 3 presents the results for electron energetics, and the finding that
within some range of pulsars parameters, 
the averaged Monte Carlo behaviour of electrons does not
coincide with a solution found with the smooth method.
It also contains the explanation of these differences 
as well as consequences for predicted efficiencies
of gamma-ray emission.
Conclusions are given in Sect.\t 4.

\section{The model}
We consider electrons accelerated 
in a longitudinal electric field induced rotationally within 
the region adjacent to the surface of a neutron star.
The particles lose their energy 
due to scattering off soft thermal X-ray photons through
magnetic inverse Compton mechanism and (marginally)
due to emission of curvature radiation.
The thermal photons are assumed to originate from a flat `thermal polar cap'
with a temperature $T\sim {\rm a\ few}\times 10^6$ K and with a radius $\rth
\simeq{\rm a\ few}\times\rpc$ where $\rpc=(2\pi\rns^3/cP)^{1/2}$ is the standard
polar cap radius as determined for an aligned rotator by the open lines of 
purely dipolar magnetic field; 
($\rns$ denotes the neutron star radius and $P$ is the
pulsar period).

Since the ICS losses are to dominate over the curvature losses,
one has to ensure that electrons do not attain extremely
high Lorentz factors ($\gamma < 10^6$). 
Therefore, either the electric field should be weak or at least the size of the accelerating
region should be small enough to prevent $\gamma \ga 10^6$. 
For this reason most papers focusing on the resonant ICS
made use of
a relatively weak electric field after Michel \cite{michel}
(for example: Sturner \cite{sturner} and recently Supper \& Tr\"umper \cite{supper}). 
Michel's model is considered as unrealistic since it ignores magnetic field line
curvature (detailed treatment of this problem was first presented 
by Arons \& Scharlemann \cite{as}) as well as inertial frame dragging 
(its importance was acknowledged and the effect was worked out for the first time 
by Muslimov \& Tsygan \cite{mt}).
Both effects lead to significantly stronger electric fields operating in polar-cap
accelerators.

However, for the sake of comparison with previous papers on
the resonant ICS
we first consider the weak electric field after Michel \cite{michel}.
Accordingly, we assume that electrons are accelerated in a constant 
$E_\parallel$ extending between stellar surface ($h = 0$)
and altitude $h_{\rm acc}$ : 
\begin{equation}
E_\parallel =\left\{           \begin{array}{l}
\left( \frac{8\pi mc \bpc}{eP} \right)^{1/2} \simeq\\ 
\hbox{\hskip1cm} \simeq 3.58\cdot 10^{5}\,
\left(\frac{B_{12}}{P}\right)^{1/2}{\rm V\ cm^{-1}} , \\*[2mm]
\hbox{\hskip3.875cm} {\rm for}\ 0 \le h \le h_{\rm acc};\\*[3mm]
\hbox{\hskip7mm} 0, \hbox{\hskip28.5mm} {\rm for}\ h > h_{\rm acc}.
\end{array}\right.
\label{elfield1}
\end{equation}
with $\hacc = \rpc$, and where $P$ is in seconds, $B_{12}$ is $\bpc$ in Teragauss. 
Fixed location of $\hacc$ means that the
model is not self-consistent in the sense that 
the accelerating field will not be shorted out by a pair formation front
(should it occur at a height below $\hacc = \rpc$).

As the second case for $E_\parallel$ we take 
the advanced model of an electric 
field in the form elaborated by Harding \& Muslimov (\cite{hm}).
HM98 were the first to incorporate electron energy losses due to the ICS 
in a strong-$E_\parallel$ acceleration model. However,
this was done to introduce a high altitude accelerator 
with negligible resonant ICS losses (an idea of an unstable ICS-induced 
pair formation fronts). In that work Harding and Muslimov also calculated self-consistent
values of the acceleration height $\hacc$ 
using an accelerating electric field which takes into
account  the upper pair formation front (eqs.~(18) and
(23) in HM98).

For $\hacc \la 0.1 \rpc$ 
the field is well approximated with 
\begin{eqnarray}
\lefteqn{E_\parallel (h) \simeq 3\ \frac{\Omega \rns}{c}\ \frac{\bpc}{1 - \epsilon_{\rm GR}}\
\frac{h}{\rns} \frac{\hacc}{\rns} \left(1 - \frac{h}{\hacc}\right)
\times}\nonumber\\
&&\hbox{\hskip9mm}\times\left[\kappa \cos{\chi} + A \sin{\chi}\right],
\label{elfield2}
\end{eqnarray}
where $\Omega = 2\pi/P$, $(1 - \epsilon_{\rm GR}) \simeq 0.6$, $\kappa \simeq
0.15$, $\chi$ is the angle between the magnetic dipole and rotation axes, 
$h$ is the altitude, and the factor
$A \sin{\chi}$ is negligible for nonorthogonal rotators (see HM98 
for details).
For $\hacc$ approaching $\rpc$
the linear dependence on $\hacc$ in Eq.(\ref{elfield2}) disappears.
We have found 
that for $0.5\rpc\la \hacc \la 3\rpc$ and $h \la \rpc/3$ 
the electric field given by eq.(18)
in HM98 is well aproximated with a formula
\begin{eqnarray}
\lefteqn{E_\parallel (h) \simeq 3\ \frac{\Omega \rns}{c}\ \frac{\bpc}{(1 -
\epsilon_{\rm GR})^{1/2}}\
\frac{h}{\rns} \frac{\rpc}{\rns} \left(1 - 
\frac{h}{\hacc}\right) \times}\nonumber\\
&&\hbox{\hskip9mm}\times\left[\kappa \cos{\chi} + A \sin{\chi}\right].
\label{elfield3}
\end{eqnarray}

For purposes of comparison as well as  to simplify our calculations,
we inject electrons from the
center of polar cap and propagate them along the straight magnetic axis 
field line.
The propagation proceeds in steps
of a size $dh$,
which is one hundred times smaller than any of characteristic length
scales involved in the problem, like the acceleration length scale
or the mean free path for the magnetic ISC.
In each of the distance steps the energy of an electron is increased by a
value $eE_\parallel dh$. For the sake of completeness,  we also 
include the energy loss due to the curvature
radiation $|\dot\gamma_{\rm cr}|mc^2dh/v$ where $\dot\gamma_{\rm cr}$
is the CR cooling rate given by

\begin{equation}
-\dot\gamma_{\rm cr} = \frac{2e^2\beta^4\gamma^4}{3mc\rho_{\rm curv}^2}
\label{crloss}
\end{equation}
and $v=c\beta$ is the electron velocity. Nevertheless, the CR is a very
inefficient cooling mechanism 
in the presence of acceleration models we consider below
and it could be neglected equally well.
When computing energy losses due to the CR
we assume artificially (after Sturner \cite{sturner}) 
that the magnetic axis has a fixed radius of curvature
$\rho_{\rm curv}=10^7$ cm,
though we keep the electrons all the time directly over the polar cap
center.

Magnetic inverse Compton scattering
has been treated with a Monte Carlo simulations.
The framework of our numerical code is based on 
an approach proposed by 
Daugherty \& Harding (\cite{dh89}). We improve their method of sampling 
the parameters of incoming photon 
and account for the Klein-Nishina
regime in an approximate way.

At each step in the electron trajectory, the optical depth for magnetic 
ICS is calculated 
as $d\tau=dh{\cal R}/c$, where ${\cal R}$ is 
a scattering rate, to decide whether a
scattering event is to occur.
The scattering rate $\cal R$ in the observer frame OF is calculated as

\begin{equation}
{\cal R} = c \int d\Omega \int d\eps\ \sigma
\left(\frac{dn_{\rm ph}}{d\eps d\Omega}\right)(1 - \beta\mu)
\label{rate}
\end{equation}
(eg.~Ho \& Epstein \cite{he}) where $\Omega=d\mu d\phi$ is the solid angle
subtended by the source of soft photons, ($\mu = \cos{\theta}$), $\sigma$ is
a total cross section (see below), and $dn_{\rm ph}/d\eps/d\Omega$ 
is the density of the soft photons per unit energy and per unit solid angle.
The symbol $\eps$ denotes photon energy in dimensional energy units.
Hereafter we will use its dimensionless counterpart $\epsilon =
\eps/(mc^2)$ to denote the photon energy in the observer frame OF and the
primed symbol
$\epr=\epsilon\gamma(1 - \beta\mu)$ in the electron rest frame ERF.

For temperatures of the thermal polar cap and Lorentz
factors of electrons considered below, photon energies $\epr$ may fall
well above $\eb=B/\bcr$, 
a local magnetic field strength in units of the critical
magnetic field $\bcr = m^2c^3(e\hbar)^{-1}$.
This
suggests a full form of the relativistic cross section for Compton scattering 
in strong magnetic fields to be used 
(Daugherty \& Harding \cite{dh86}).
However, incoming photons that propagate in the OF at an angle $\theta = \cos^{-1}{\mu}$
are strongly collimated in the ERF with a cosine of polar angle 
$\mu^\prime = (\mu - \beta)/(1 - \beta\mu)$ close to $-1$.
As was shown by Daugherty \& Harding (\cite{dh86}), when $|\mu^\prime|$
approaches $1$, resonances at
higher harmonics become narrower and weaker, and scattering into higher
Landau
states becomes less important. In such conditions, the polarization-averaged
relativistic magnetic
cross section in the Thomson regime is reasonably well approximated with a
nonrelativistic, classical limit:

\begin{equation}
\sigma = \frac{\st}{2}\left(1 - {\mu^\prime}^2 + (1 +{\mu^\prime}^2)
\left[g_1 + \frac{g_2 - g_1}{2}\right]\right)
\label{crosssection}
\end{equation}  
where $\st$ is the Thomson cross section, 
and $g_1$ and $g_2$ are given by

\begin{equation}
g_1(u) = \frac{u^2}{(u+1)^2}, \hbox{\hskip 1cm}
g_2(u) = \frac{u^2}{(u-1)^2 + a^2}
\end{equation}
with $u\equiv\epr/\eb$, $a\equiv 2 \alpha_f\eb/3$, where $\alpha_f$ is a
fine-structure constant (eg.~Herold \cite{herold}; Dermer \cite{dermer}).

In the Klein-Nishina regime ($\epr \ga 1$) the relativistic magnetic cross 
section for the $|\mu^\prime|\approx 1$ case
becomes better approximated with the well
known Klein-Nishina relativistic nonmagnetic total cross section $\skn$
(Daugherty \& Harding \cite{dh86}; Dermer \cite{dermer}).
When $\epr \gg \eb$, (the condition fulfilled in the K-N regime since 
$\eb < 1$ holds throughout this paper), 
the resonant term $(g_2 - g_1)/2$ in
Eq.\t(\ref{crosssection}) becomes negligible 
and the nonresonant term $g_1$ approaches unity which results in
$\sigma\approx\st$.
Therefore, to approximate the Klein-Nishina decline
we replace the single nonresonant term $g_1$ in the square
bracket of Eq.\t(\ref{crosssection}) with $\skn/\st$ for $\epr > 2\eb$.

To make calculations
less time-consuming, we have used the delta-function
approximation for the resonant part of the cross section (Dermer \cite{dermer}),
when calculating the optical depth for a scattering and the averaged 
energy loss of electron (Eq.\t \ref{lossrate}, see below).

As the density of soft photons $dn_{\rm ph}/d\eps/d\Omega$ we take the spectral
density of blackbody radiation 
given by

\begin{equation}
\left(\frac{dn_{\rm ph}}{d\eps d\Omega}\right)d\eps\, d\Omega
= \frac{2}{\lambda_{\scriptscriptstyle C}^3}
\frac{\epsilon^2\, d\epsilon\, d\Omega}{[\exp(\epsilon/{\cal T}) - 1]}
\end{equation}
where 
${\cal T}$ is a dimensionless temperature (${\cal T}\equiv \frac{kT}{mc^2}$)
and $\lambda_{\scriptscriptstyle C}=h/(mc)$ is the electron Compton
wavelength. Hereby we neglect anisotropy of the thermal emission expected in
the strong magnetic field (eg Pavlov et al.~\cite{psvz}). 
For $\epsilon \ll \eb$
a preferred photon propagation direction is that of the magnetic field 
because the opacity is reduced for radiation polarized across $\vec{B}$.
The probability of collision with an electron is greatly reduced for
photons propagating at $\mu \approx 1$,
thus,
the efficiency of ICS as calculated below should be treated as an upper limit.

If a scattering event occurs, we draw the energy $\epsilon$
and the cosine of polar angle $\mu$ for an incoming photon 
from a distribution determined with the
integrand of Eq.\t(\ref{rate}).
We do this with the simple two-dimensional rejection method which 
is reliable though
more time-consuming (cf. an approach by Daugherty \& Harding \cite{dh89}).
Next, we transform $\epsilon$ and $\mu$ to the ERF values $\epr$ and
$\mu^\prime$, and sample
the cosine of polar angle of an outgoing photon $\mu^\prime_s$ using 
the differential form of the cross section (\ref{crosssection}) in the Thomson
limit:

\begin{eqnarray}
\lefteqn{\frac{d\sigma}{d\epr_s d\Omega^\prime_s} = \frac{3\st}{16\pi}
\,\delta(\epr_s - \epr_{\rm scat})
\left[
(1 - {\mu^\prime}^2)(1 - {\mu^\prime_s}^2)\ + \right.}\nonumber\\
&&\hbox{\hskip12mm}\left.+\ \frac{1}{4}\ (1 + {\mu^\prime}^2)
(1 + {\mu^\prime_s}^2)(g_1 + g_2)\right]
\label{diffsection}
\end{eqnarray}
(eg.~Herold \cite{herold}), where $d\Omega^\prime_s = d\phi^\prime_s d\mu^\prime_s$ is 
an increment of solid angle into which outgoing
photons with energy $\epr_s$ in the ERF are directed. 
As in the case of
total cross section (\ref{crosssection}), when $\epr >
2\eb$ we replace the factor $(g_1 + g_2)$ in (\ref{diffsection}) with
$(2\skn/\st + g_2 - g_1)$.
The sampled value of $\mu^\prime_s$ determines 
the energy $\epr_{\rm scat}$ for the particular scattered photon 
with the relativistic formula: 

\begin{eqnarray}
\lefteqn{ \epr_{\rm scat} = \left(1 -
{\mu^\prime_s}^2\right)^{-1}\left\{\vbox{\vskip5mm} 
1 + \epr(1 - \mu^\prime\mu^\prime_s)\ + \right.}\nonumber\\
&&\hbox{\hskip4mm}\left.-\left[  1 + 2\epr\mu^\prime_s (\mu^\prime_s -
\mu^\prime) +
{\epr}^2 (\mu^\prime_s - \mu^\prime)^2  \right]^{1/2}\right\}
\end{eqnarray}
appropriate for collisions with a recoiled electron remaining at the ground Landau
level (Herold \cite{herold}). Finally, a value of the electron's longitudinal momentum 
in the ERF is changed due to recoil from zero to $(\epr\mu^\prime -
\epr_{\rm scat}\mu^\prime_s)mc$, 
and transformed back to the OF.

The Monte Carlo method will be compared with a 
smooth integration treatment
of the magnetic ICS (Chang \cite{chang}; Sturner
\cite{sturner}).
In the integration method we assume the same procedure to accelerate
an electron and to subtract its energy losses due to the curvature radiation
as described above (Eq.\t \ref{crloss}).
The only difference is in accounting for electron energy losses
due to the magnetic inverse Compton scattering. 
These are estimated in \emph{each} step (regardless the value of 
the optical depth for the
scattering process) from the following formula for the mean electron
energy loss rate:

\begin{eqnarray}
\lefteqn{-\dot{\gamma}_{\sss \rm ICS} = c\int
d\epsilon \int d\Omega\, \left(\frac{dn_{\rm ph}}{d\epsilon d\Omega}\right)
(1 - \beta\mu)\ \times}\nonumber\\
&&\hbox{\hskip20mm}\times \int
 d\epr_s \int
d\Omega^\prime_s
\left(  \frac{d\sigma}{d\epr_s d\Omega^\prime_s}  \right)
(\epsilon_s - \epsilon)
\label{lossrate}
\end{eqnarray}
where 
$\epsilon_s = \epr_s\gamma(1 + \beta\mu^\prime_s)$ is the scattered
photon energy in the OF (eg.~Dermer \cite{dermer}). In other words,
the electron Lorentz factor is determined as a solution of the differential
equation:

\begin{equation}
\frac{d\gamma}{dh} = v^{-1}(\dot\gamma_{\rm acc} + 
\dot\gamma_{\sss \rm ICS} + \dot\gamma_{\rm cr}) .
\label{diffeq}
\end{equation}

At each step, 
as an electron moves upwards,
a decrease in both the dipolar magnetic field strength and
the solid angle subtended by the thermal polar cap is taken
into account in both methods.

\section{Results}

We compare the Monte Carlo method described in Sect.\t 2
with the integration method (numerical integration of
Eq.\t (\ref{diffeq})), for a pulsar with the polar magnetic field 
strength and rotation period as for \object{B1509$-$58}
($\bpc=15.8 \times 10^{12}$ G,  $P=0.15$ s).

\subsection{The case of the M74 electric field }

\begin{figure*}   
\resizebox{12cm}{!}{\includegraphics{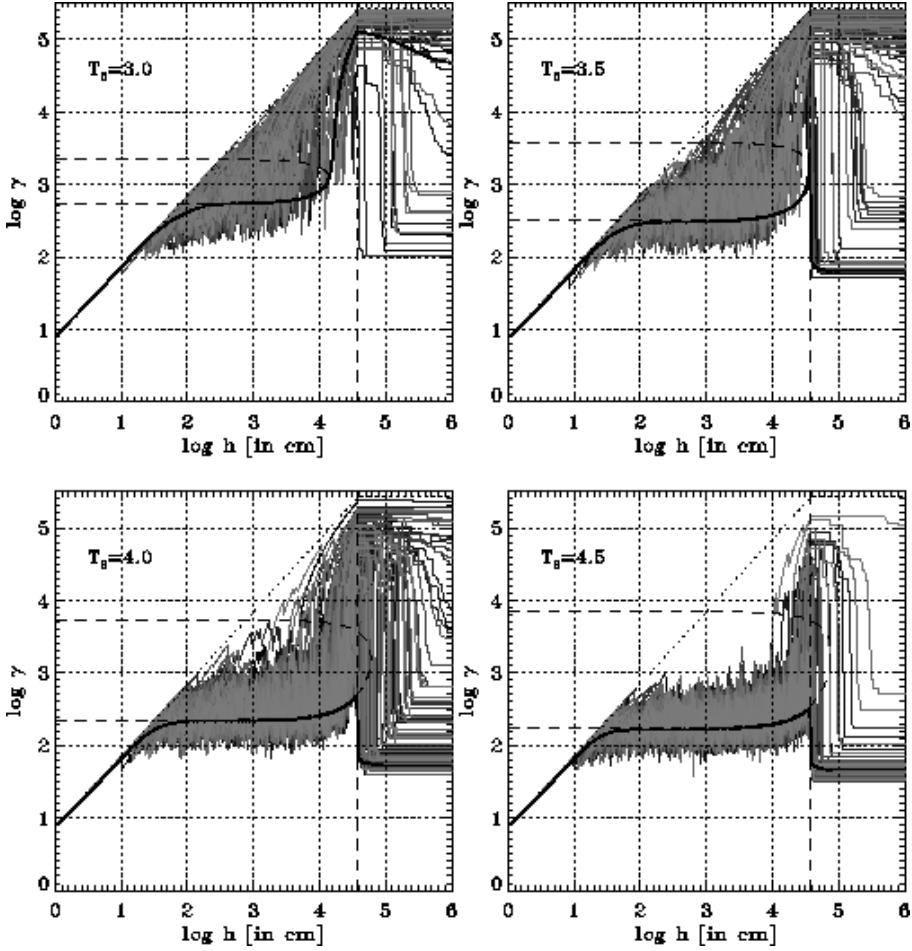}}
\hfill
\parbox[b]{55mm}{
\caption{Electron Lorentz factor as a function of altitude
for the M74 electric field (Eq.\ref{elfield1})
above the surface 
of a neutron
star 
with $B=15.8\times 10^{12}$ G, $P=0.15$ s and $\rth=10^5$ cm for 
four temperatures of thermal polar cap: $T_6=3.0$, $3.5$, $4.0$, and $4.5$.
Each panel represents 100 curves calculated for individual electrons 
in the Monte Carlo way (grey). 
The mean behaviour determined separately by solving numerically the
differential equation (\ref{diffeq}) 
is overplotted on them as the thick solid line. 
A strong disagreement between the two methods stands out for $T_6=3.5$.
The dotted line is for the
case with no radiative losses.
The thick dashed line surrounds a region with $\scloss > \gain$.
The upper boundary of acceleration region for model of Michel (\cite{michel}) 
is marked by the vertical dashed line.}
\label{fig1}}
\end{figure*}

In this subsection the results are presented for 
the electric field  
of Michel (\cite{michel}) (see Eq.\ref{elfield1}).
The assumed values of $\bpc$ and $P$ give $E_{\parallel} = 3.67\times 10^6$ V cm$^{-1}$
and the standard polar cap radius $\rpc=3.74\times 10^4$ cm.
Following Sturner (\cite{sturner}), we assumed the thermal polar cap radius $\rth=10^5$ cm 
($\simeq 2.7\times\rpc$) and $\hacc = \rpc$.
Calculations have been performed for 
four different temperatures of the thermal polar cap: 
$T_6 = 3.0$, $3.5$, $4.0$ and $4.5$, where $T_6=T/(10^6 {\rm K})$.

Each panel in Fig.~\ref{fig1} presents in grey 
the curves $\gamma(h)$ as calculated in the Monte Carlo way for one 
hundred individual electrons.
The behaviour determined with Eq.\t (\ref{diffeq}) 
 is overplotted as the
thick solid line. 
 The inclined dotted line presents changes of electron energy if
there were no losses. It becomes horizontal at altitude 
$h_{\rm acc}=\rpc=3.74\times10^4$ cm above which 
the longitudinal component of electric
field $E_\parallel$ is assumed to be screened by a charge-separated 
plasma distribution (see Eq.\t \ref{elfield1}). This
height is denoted by the vertical long dashed line.

\begin{figure}
\resizebox{\hsize}{!}{\includegraphics{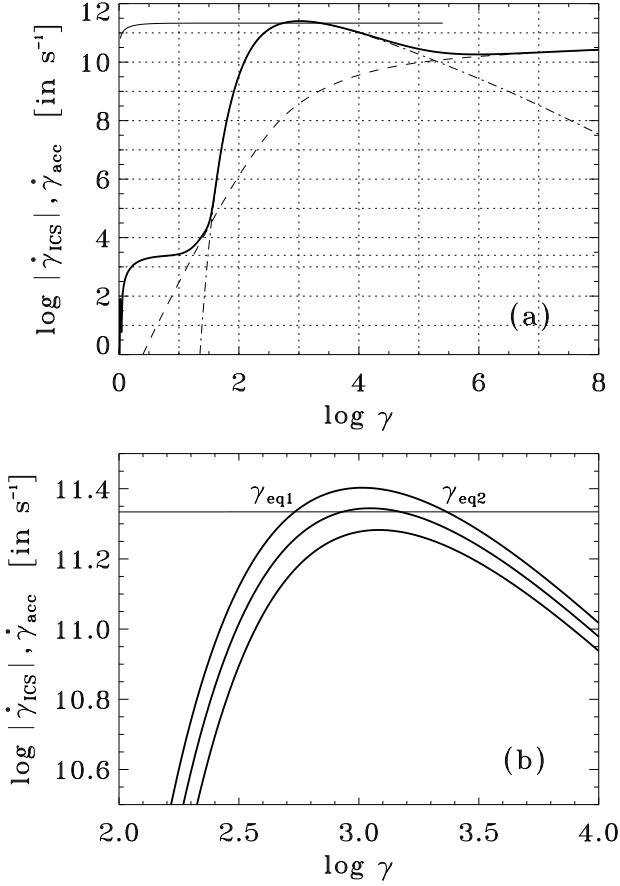}}
\caption{Electron energy loss rate due to the ICS (thick solid line)
and the acceleration rate (thin solid)
as a function of the electron Lorentz factor for the case $T_6=3.0$: 
{\bf a} case $h=0$ with contributions from angular, nonresonant, and resonant part
of the cross section shown explicitly (dotted, dashed, and dot-dashed line,
respectively); {\bf b} 
decrease in the loss rate with increasing altitude. The curves are for
$h=0$, $10^4$, and $2\times 10^4$ cm (from top to bottom).
Note the narrowness of the Lorentz factor range within which $\scloss$ exceeds
$\gain$.}
\label{fig2}
\end{figure}

The thick solid trajectories $\gamma(h)$ in four panels of Fig.~\ref{fig1}
represent the "averaged
treatment" solutions and
are in good agreement with the results
of previous calculations by Sturner (\cite{sturner}) (cf his Fig.~4).
Below $h\sim 10$ cm they overlap with the case with no
energy losses (dotted), since the acceleration rate of electron (as given by
$\dot\gamma_{\rm acc}=\frac{eE_\parallel}{mc^2}\beta c$) significantly 
exceeds 
the energy loss rate due to the ICS (see Fig.~\ref{fig2}a).
Energy losses due to the CR are negligible for any Lorentz factor
accessible for an electron, given the assumed acceleration model 
of Michel (Eq.~\ref{elfield1}).
At larger altitudes
the thick trajectories illustrate the acceleration stopping effect 
noted by Chang (\cite{chang}) and Kardash\"ev et al.~(\cite{kmn}):
the energy that the electrons would have gained 
between altitudes $h\sim10^2$ and $\sim 10^4$ cm
due to the electric field $E_\parallel$
is transferred to the thermal photons through the resonant inverse Compton
scatterings.
Depending on which process -- either the resonant ICS cooling or the acceleration -- 
ceases first, the electrons end up with energy which is either closer to
$e E_\parallel h_{\rm acc}$ (ie.~the energy acquired  
due to the full voltage drop at no radiative losses, the case $T_6=3.0$) or
decreases towards a few tens $\times mc^2$ (the cases with $T_6\ge 3.5$). 

As can be seen in Fig.~\ref{fig1}, the Monte Carlo tracks generally
behave qualitatively in the same way: the increasing temperature of the
thermal cap increases the altitude at which the acceleration takes over,
and eventually, the Lorentz factors of most of  electrons become limited 
below $10^2$.
However,
there are strong differences  between the average energy of the Monte Carlo 
electrons and the value obtained with the integration method. 
They are especially pronounced for $T_6=3.5$, which is the case 
where a maximum altitude at which the acceleration 
can be counterbalanced by 
the electron energy losses due to the resonant ICS is equal to $h_{\rm acc}$.
The bulk of the Monte Carlo tracks ends up with 
Lorentz factors $\gamma > 10^5$, whereas the integration-method solution
gives $\gamma \simeq 64$.
To understand these differences it is worth investigating closely
the behaviour of one exemplary electron as determined by the two methods.

In the integration method, the electron's energy increases monotonically
until an equilibrium settles between the energy
loss process and the acceleration.
Fig.~\ref{fig2}b shows $\scloss$ and $\dot\gamma_{\rm acc}$ as a function of
the Lorentz factor $\gamma$ for three different altitudes in the case 
$T_6=3.0$. One can see that $\scloss$ exceeds $\dot\gamma_{\rm acc}$
between the two equilibrium Lorentz factors $\geql$ and $\geqh$ 
the values of which are determined by the condition 
$\scloss = \dot\gamma_{\rm acc}$.
We have calculated them using the approximation $\dot\gamma_{\sss\rm ICS} 
\simeq \dot\gamma_{\rm res}$ 
where $\dot\gamma_{\rm res}$ is the energy loss rate
due to the resonant part of the cross section alone (cf.\t Eq.\t(22) in 
Sturner \cite{sturner}).
 The two equilibrium Lorentz factors $\geql$ and $\geqh$
as calculated for different altitudes $h$
are presented in Fig.~\ref{fig1} by the thick dashed line.
For increasing altitude their values
approach each other because a decrease in the density of black-body 
photons moves down
the whole curve $|\dot\gamma_{\sss\rm ICS}(\gamma)|$ in Fig.~\ref{fig2}b.
Thus, the energy-altitude space is divided into two regimes:
the region
where $\scloss > \gain$, hereafter called the "energy loss
dominated" region, (it is surrounded by the thick dashed line 
in Fig.~\ref{fig1}),
and the "acceleration dominated" region where $\scloss < \gain$.
As can be seen in Fig.~\ref{fig1}, according to the integration method
the trajectory of an
electron in the energy-altitude space 
does not penetrate the energy loss dominated region.
As the electron moves upwards, its Lorentz factor settles at 
the lower equilibrium value $\geql$, and stays there as long as 
the equilibrium can exist ie.\t until the curves
$|\dot\gamma(\gamma)|$ and $\dot\gamma_{\rm acc}(\gamma)$ in Fig.~\ref{fig2}b
disconnect. For $T_6=3.0$ this happens near the altitude $\heq=2\times10^4$.
Above the height $\heq$
the acceleration cannot be counterballanced by the ICS energy losses 
at any Lorentz factor accessible for the electron 
and the increase in the electron energy resumes.

\begin{figure}
\resizebox{\hsize}{!}{\includegraphics{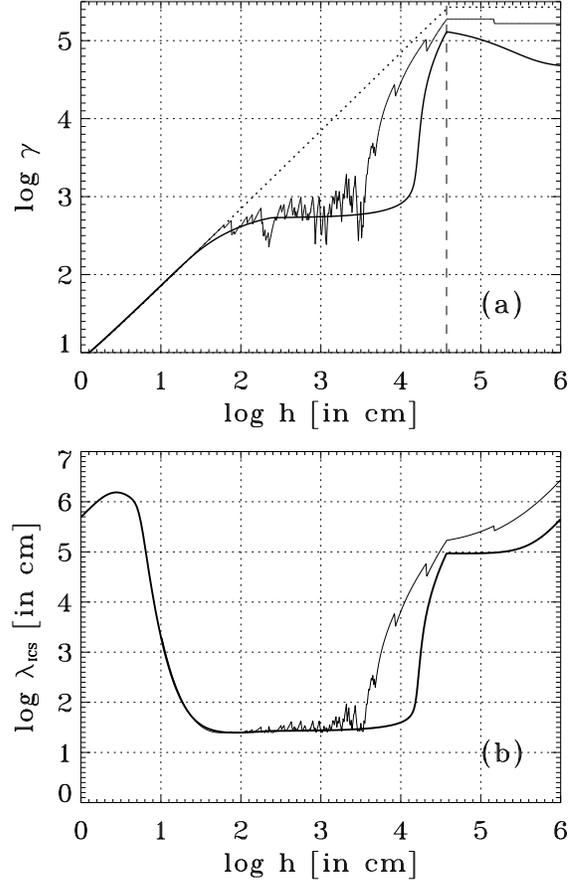}}
\caption{Lorentz factor of an electron ({\bf a}) 
and the corresponding local mean free path for
the magnetic Compton scattering ({\bf b}) as a function of altitude for the case
$T_6=3.0$.
The erratic line is an example of the Monte Carlo trajectory.
The thick solid line represents a solution of the differential equation
(\ref{diffeq}).
Note a large range of $\gamma$ for the Monte Carlo track between
$h=10^2$ and $3\times 10^3$ cm.}
\label{fig3}
\end{figure}

In the Monte Carlo method, the electron's energy losses due to the
scattering occur in a discontinuous way, thus, $\gamma$ oscillates
around the value $\geql$ (see Fig.~\ref{fig3}).
When the energy loss rate $\scloss$ hardly exceeds the acceleration rate 
(see Fig.~\ref{fig2}) the 
difference between the equilibrium Lorentz factors $(\geql - \geqh)$
is small and the electron is able to pass \emph{through}  
the energy-loss dominated region.
Eg.\t for the case $T_6=3.0$, $\geqh\simeq 2200\simeq 4\times \geql$ 
at $h=10^3$ cm.
To increase its energy from $\geql$ to $\geqh$
 it is enough for the electron to avoid a scattering 
over a `break-through'-distance 
$\Delta_{\rm bt} = \frac{\geqh -\geql}{eE_\parallel/mc^2}=228\, {\rm cm}$
which is only `a few' times larger than 
the local mean free path for scattering $\lambda_{\sss \rm ICS}$.
The local value of $\lambda_{\sss \rm ICS}$ depends strongly on the electron
Lorentz factor and for $\gamma$
between $\geql$ and $\geqh$ it
ranges from $28$ to $108$ cm (the case $T_6=3.0$, $h=10^3$ cm). 
This gives $\Delta_{\rm bt} 
\simeq (8 - 2)\times \lambda_{\sss \rm ICS}$.
Thus, there is a very large probability for the electron to gain 
$\gamma \ga \geqh$ at an altitude $\hmc$ lower than $\heq$.
Once the electron 
enters the acceleration dominated region at $\hmc \ll \heq$,
its energy starts to increase up to a
value much larger than obtained in the integration method.

\begin{figure}
\resizebox{\hsize}{!}{\includegraphics{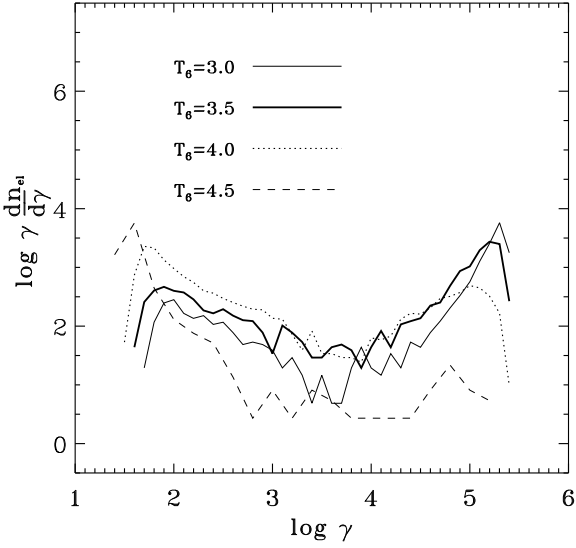}}
\caption{Energy distributions of electrons
at $h=10^6$ cm 
for $T_6=3.0$, $3.5$, $4.0$, and $4.5$. 
The energy unit is $mc^2$ and the convention analogous to $F_\nu$ is used.
Both the total power of a single distribution 
and its hardness decrease with increasing $T$.}
\label{fig4}
\end{figure}

For increasing temperatures $T$, the width $(\geqh - \geql)$ 
of the energy-loss dominated region increases (Fig.~\ref{fig1})
whereas the mean free path for $\gamma=\geql$  decreases.
This makes the diffusion of electrons through the energy loss dominated
region more difficult: to increase its energy from $\geql$ to $\geqh$ an
electron must avoid a scattering over a distance which is increasing
multiplicity of the local mean free path.
As a result, the energy distribution for outgoing electrons becomes softer
(Fig.~\ref{fig4}).

\begin{figure}
\resizebox{\hsize}{!}{\includegraphics{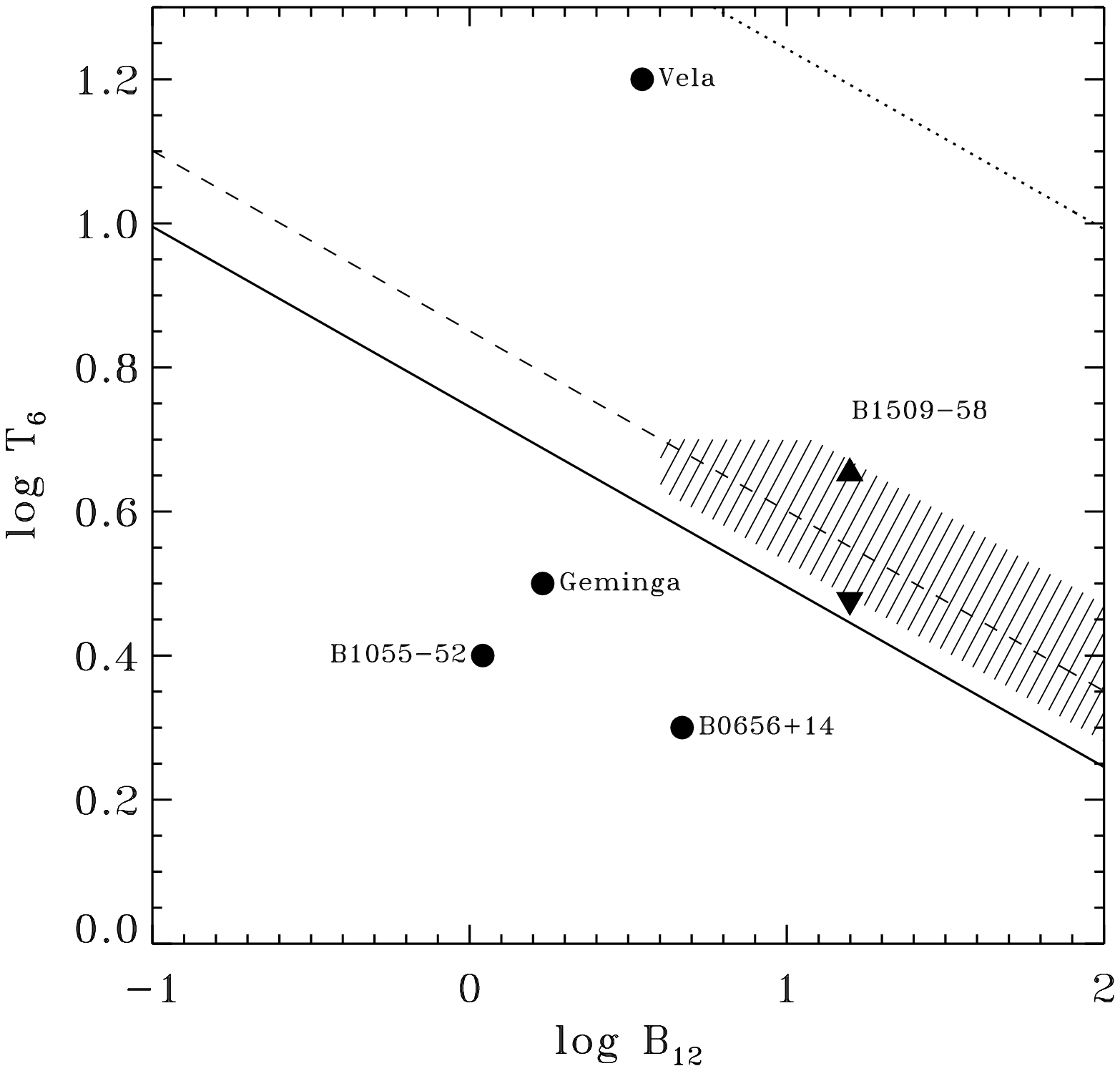}}
\caption{Approximate region of strong discrepancies between results
of the Monte Carlo
and the integration method for the M74 electric field case 
(Eq.\ref{elfield1}),
for pulsars with $P\simeq 0.15$ s and $\rth >
\rpc$ (hatched).
The solid line 
marks the equilibrium between the value of the energy loss rate
$\scloss$ near the resonance and the acceleration rate $\gain$ at the
surface level (Eq.\t \ref{eqcon2} with $h=0$).
The dashed line indicates the equilibrium at $h=\rpc$ for $\rth=10^5$ cm
and the dotted line is for $h=\rpc$, $\rth=10^4$ cm.
Dots are positions of four pulsars with putative thermal 
X-ray emission (\"Ogelman \cite{ogelman}). The temperature of a thermal cap
is not known for \object{B1509$-$58}: triangles mark the range of $T$
considered in Fig.~\ref{fig1}.}
\label{fig5} 
\end{figure}

It should be emphasized that the strong disagreement between 
the final electron energies as determined with the 
two methods only appears if conditions similar to those for the case
$T_6=3.5$ in Fig.~\ref{fig1} are fulfilled.
These include the equilibrium between
the maximum rate of resonant energy losses 
and the rate of acceleration at $h=h_{\rm acc}$.
Making use of Dermer's approximation for $\resloss$ (Dermer \cite{dermer})
one can easily find that 
it  has the maximum at $\gamma_{\rm res}=\eb/(w(1 - \beta\mu_{\rm min}){\cal T})$
where $\mu_{\rm min} = h/(h^2 + \rth^2)^{1/2}$ is a cosine of an angle 
at which the thermal cap radius is
seen from the position of electron and $w=-\ln{0.5}$.
The equilibrium $\resloss = \gain$ holds for

\begin{equation}
T_6^{\rm eq} = 4.14\left(\frac{\beta^2\ \frac{eE_\parallel}{mc^2}}{(1 -
\beta\mu_{\rm min})\ B_{12}(h)}\right)^{1/2}
\label{eqcon1}
\end{equation}
where $B_{12}(h)=B(h)/(10^{12} {\rm G})$ and
$\frac{eE_\parallel}{mc^2}$ is in cm$^{-1}$ (cf Eq.\t
(29) in Chang \cite{chang}).
For the considered accelerating field (Eq.\t \ref{elfield1})
this condition becomes

\begin{equation}
T_6^{\rm eq} = 3.46\ \beta\ (1 - \beta\mu_{\rm min})^{-1/2}
\left(B_{12} P\right)^{-1/4}
\label{eqcon2}
\end{equation}
and is shown in Fig.~\ref{fig5} for $h=0$ and $P=0.15$ s as the solid line.

Other conditions required for the diffusion 
are low thermal cap temperatures and high magnetic field strengths.
Both requirements partially stem from the dependence 
$\mfp(\gamma_{\rm res})\propto 
B T^{-3}$ which results in $\mfp \ll \Delta_{\rm bt}$ for 
high $T$ and low $\bpc$. [Note that at the resonance $\mfp$  is
\emph{proportional to} $B$ just as $\resloss$]. 

Additionally, high temperatures and low $B$-field strengths preclude the
diffusion because of relatively large energy losses due to scatterings
in the Klein-Nishina regime (see Fig 2a).
First, for $\gamma > \gamma_{\rm res}$, 
$\knloss \ll \gain$ must hold, which requires low $T$.
Second, the resonant bump
in the $|\dot\gamma_{\sss \rm ICS}(\gamma)|$ curve
must be present, which is possible when $|\dot\gamma_{\rm res}(\gamma_{\rm
res})| \gg |\dot\gamma_{\sss \rm KN}|$.
Since $\dot\gamma_{\rm res}(\gamma_{\rm res})/\dot\gamma_{\sss \rm KN} 
\propto B$, strong magnetic fields are preferred.

For $P\simeq 0.15$ s and $\rth > h_{\rm acc}$ all these
constraints
limit the diffusion regime to $T\la 5\times 10^6$ K and $\bpc \ga 3\times
10^{12}$ G with the $T$ and $\bpc$ roughly fulfilling Eq.\t (\ref{eqcon2})
taken for $h=\hacc$. The region of strong discrepancies between the two
methods is shown in Fig.~\ref{fig5} (hatched).

Among a few pulsars exhibiting a two-component thermal X-ray spectra
only the \object{Vela} pulsar has an inferred temperature of the thermal polar cap 
($T\simeq 15.8\times 10^6$ K, \"Ogelman \cite{ogelman})
which is sufficient for the acceleration to be initially halted by the resonant ICS
(ie.~it exceeds $T_6^{\rm eq}$ for $h\ll\rth$, see Fig.~\ref{fig5}).
The equilibrium $|\dot\gamma_{\rm res}|=\gain$ 
holds up to the altitude $\heq\simeq 2.2\times 10^4$ cm which is two times
larger than the inferred thermal cap radius ($10^4$ cm, Sturner
\cite{sturner}) and
two times smaller than $\rpc=4.8\times 10^4$ cm. 
Nevertheless, because of the high $T$, 
at $h\ll \heq$ the diffusion is precluded by the 
extremely small $\mfp\sim 0.1$ cm. It becomes efficient only at $h\ga
0.9\heq$, thus,
most Monte-Carlo electrons follow closely the behaviour predicted by 
the continuous approach by Chang (\cite{chang}) or Sturner (\cite{sturner}).

For other pulsars with a likely thermal X-ray emission from heated
polar caps (eg.~\object{Geminga},
\object{B1055$-$52}, and \object{B0656$+$14}) inferred temperatures are much lower 
(between 2 and  4
$\times 10^6$ K). Therefore, $\scloss \ll \gain$ for any Lorentz factor
accessible for an electron (Fig.~\ref{fig5}) and the Compton scattering cannot 
considerably redistribute
an initial electron energy spectrum.

Both, in the case of negligible ICS losses, ie.~for $T \ll T^{\rm eq}(h=0)$, 
and in the cases when 
$|\dot\gamma_{\rm res}(\gamma_{\rm res})| \gg \gain$ holds up to $h=h_{\rm
acc}$, ie.~for $T\gg T^{\rm eq}(h=h_{\rm acc})$,
we find the energy distribution of electrons to be quasi-monoenergetic 
with energy well approximated by Eq.\ref{diffeq}.
In the case of negligible ICS cooling the energy distribution of electrons 
is a narrow peak at the energy $\gamma_{\rm max} mc^2 = 
e E_\parallel h_{\rm acc}$, whereas
for very large efficiency of ICS losses (close to 100\% of $\gamma_{\rm
max}mc^2$)
nearly all particles are cooled down to 
the energy $\gamma_{\rm min}$ for which the ICS loss rate
decreases sharply (see Fig.~\ref{fig2}a and Fig.~\ref{fig1}).
Between these limiting cases, ie. when the ICS cooling is comparable to the
acceleration, broad energy distributions of electrons (ranging from
$\gamma_{\rm min}$ to $\gamma_{\rm max}$) emerge (Fig.~\ref{fig4}).

\subsection{The case of the HM98 electric field}

To enable electron energy losses due to the resonant ICS to compete with
acceleration (the resonant ICS damping becomes then important
and the diffusion effect does occur) the strength of the electric field $E_\parallel$
should not exceed a critical value which may be roughly estimated 
as $E_\parallel^{\rm eq} \simeq 3 \cdot 10^4 B_{12} 
T_6^2\, {\rm V \, cm^{-1}}$ (cf.~Chang \cite{chang}).
For $E_\parallel (h)$ given by eq.~(18) of HM98
this may only occur in the case of small acceleration length $\hacc \ll \rpc$
(see Eq.~(\ref{elfield2})), when
the acceleration height is limited by the ICS-induced pair
formation front. Self consistent values of $\hacc$ are then of the order of 
a few$\times 10^3$ cm, i.e. much smaller than for a CR-induced case
(cf.~figs.~5 and 6 in HM98). A value of $\hacc$ is considered self consistent if
the acceleration of primary electrons by the electric field induced with this $\hacc$
generates pair formation front at $\hacc$. 

We calculated  self consistent $\hacc$ in an approximate way, following HM98,
by finding a minimum of 
a sum of acceleration length and a mean free path for one photon
absorption: 

\begin{equation}
\hacc = {\rm min}\left\{ \left(\frac{2 \gamma}{N}\right)^{1/2}  + 
                         \left( \frac{0.2 \rho_{\rm curv}}{\eb \epsilon} \right)
                      \right\}
\label{hacc}
\end{equation}
where to get the first term we used $E_\parallel$ in the limit $h \ll \hacc$, given by
Eq.~(\ref{elfield2}): $E_\parallel (h) = N h$ with
$N=3\Omega\rns\bpc\kappa\cos{\chi}(c(1 - \epsilon_{\rm
GR})\rns^2)^{-1} \hacc$.
As an energy of pair producing photons in the second term we
assumed that for the resonant scattering: $\epsilon \sim \gamma \eb$.
In Eq.~(\ref{hacc}) we neglected a contribution from the mean free path 
for the resonant ICS which is of the order of  
$150 (1 - \beta \mu_{\rm min})^{-1} B_{12} T_6^{-3}$ cm near the resonance. 
Since in the regime $\hacc \ll \rpc$ 
the electric field $E_\parallel$ depends itself on $\hacc$ 
we had to repeat the calculation of
$\hacc$ starting from some guess value until convergence.
For $\bpc=15.8$ TG, $P=0.15$ s, $\rho_{\rm curv}=10^7$ cm and $\chi=0.1$ rad
we obtained $\hacc \simeq 2\times 10^3$ cm.

\begin{figure}
\resizebox{\hsize}{!}{\includegraphics{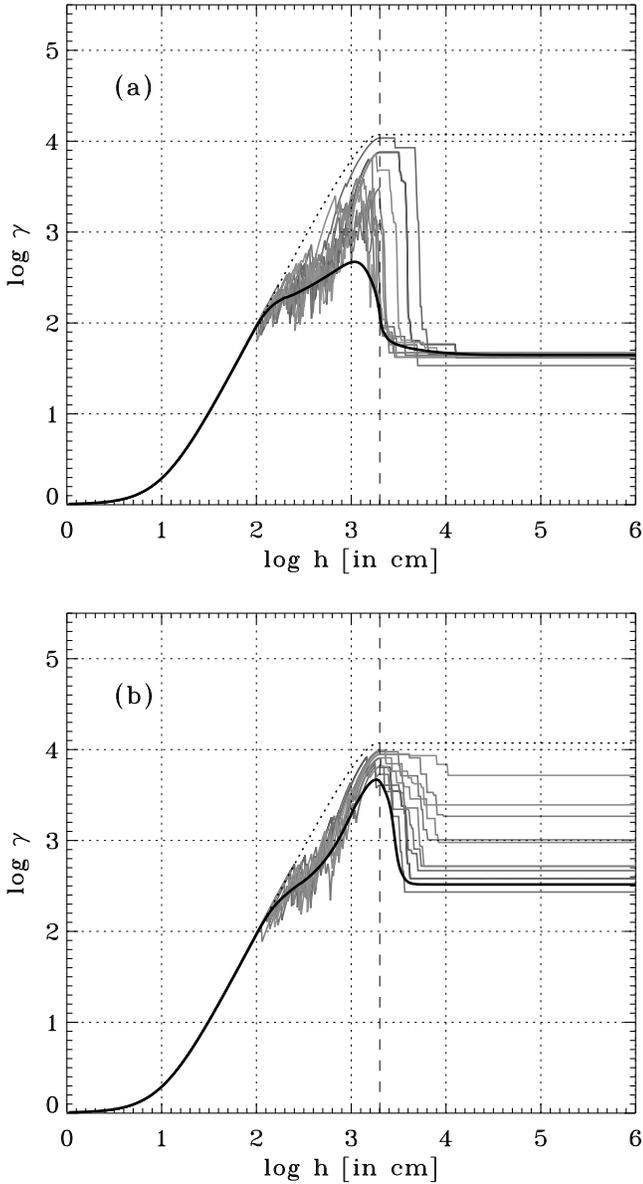}}
\caption{Electron Lorentz factor as a function of altitude
for the HM98 electric field (Eq.\ref{elfield2})
with 
$\hacc = 2\times 10^3$ cm. 
Solution found with the integration method
(thick solid line) is overplotted on ten Monte Carlo tracks.
Thick dotted line is for the case with no radiative damping.
The dashed vertical marks $h=\hacc$.
{\bf a} case of $\rth=10^5$ cm and $T_6=3.5$; 
{\bf b} case of $\rth=4\times 10^3$ cm and $T_6=3.0$;
the other pulsar parameters are the same as in Fig.~\ref{fig1}.
}
\label{fig6}
\end{figure}

Then we calculated the changes of $\gamma$ with $h$ for $\rth=10^5$ cm and  
for different thermal cap
temperatures (as listed in the preceding subsection). We found that the 
most notable divergence between
the Monte Carlo results and the integration method results occurs for $T_6 = 3.5$.
As can be seen in Fig.~\ref{fig6}a, at $h\simeq 10^3$ cm most Monte Carlo electrons 
reach Lorentz factor values between $10^3$ and $10^4$, contrary to $470$
as predicted by the integration method. The difference disappears only above the
accelerator (i.e. for $h>\hacc$) because such altitude is still low enough 
(i.e. $h\ll \rth$) 
for the resonant ICS to damp $\gamma$ to its final value of about $\sim 50$ (at which
it becomes eventually negligible).

An effect similar to that shown in Fig.~\ref{fig1}  
appears, however, when
the resonant ICS losses are no longer effective above $h=\hacc$.
Fig.~\ref{fig6}b shows the case with $\rth = 4.0\times 10^3$ cm $= 2 \times \hacc$ 
and $T_6 = 3.0$. 
Broad electron energy distribution (with average energy much
exceeding a value from the integration method) emerges.

Similarly as in the weak electric field case,
the position of the diffusion region in $T$-$\bpc$ plane (for a fixed value of $\rth$)
is here determined with a condition analogous to Eq. (\ref{eqcon1}) 
(but $E_\parallel$ of Eq.(\ref{elfield2}) requires a more accurate determination of $\hacc$)
along with a requirement 
for strong $\bpc$ and low $T$.


For stronger electric fields ($E_\parallel > E_\parallel^{\rm eq}$),
or for larger acceleration lengths the resonant ICS has negligible influence
on final electron energies.
In the first case acceleration greatly dominates the ICS damping; in the
second case acceleration is stopped only over a negligible, initial part of
accelerator height.

\section{Conclusions}

We have calculated the efficiency of energy losses for electrons
accelerated over a hot polar cap of a neutron star.
As cooling mechanisms we have considered 
the inverse Compton scattering and the curvature radiation, 
(though the latter has been negligible for the considered 
acceleration models).

The electron energy losses due to the ICS have been 
calculated with the Monte Carlo method and compared with the integration 
method (after Chang \cite{chang}; Sturner \cite{sturner}) based on 
the prescription for the averaged ICS cooling (Dermer \cite{dermer}).
We confirm general predictions of the integration approach
(eg.\t the limitation of electron Lorentz factors below $\sim 10^2$
for high temperatures), nevertheless, the ``stopping acceleration effect"
occurs at a slightly higher temperature 
than that one resulting from integrating the differential equation
(\ref{diffeq}).

For the considered pulsar parameters ($\bpc = 15.8\times
10^{12}$ G, $\rth=10^5$ cm) and the acceleration potential by Michel
(\cite{michel}), the Monte Carlo-based value is approximately 
equal to
$T_{\sss \rm MC}=4.5\times 10^6$ K (at this $T$ 
only a few percent of electrons reach $\gamma > 10^2$ at $h=10^6$ cm) 
in comparison with $T_{\rm int}=3.3\times 10^6$ as given by the
integration approach.
Although the temperature difference is small,
the final electron Lorentz factors as determined by the two methods 
may easily differ by a few orders of magnitude (Fig.~\ref{fig1}, the case
$T_6=3.5$) if the observed temperature matches this range.
Moreover, it should be kept in mind 
that for most objects with hard X-ray emission 
identified putatively as originating from the hot polar cap, 
temperatures lie within the similar range $(2 - 4) \times 10^6$ K 
(\"Ogelman \cite{ogelman}).

We find that for high-$B$ and low-$T$ pulsars 
 the integration method 
gives a poor estimation of electron Lorentz factors if 
the energy loss rate due to the resonant ICS hardly exceeds the rate of
acceleration. In the case of Michel's model and 
for $\rth > h_{\rm acc}$, this occurs if
$T\sim 4(B_{12}P)^{-1/4}$ which may be the case for \object{PSR B1509$-$58}.
In the cases when the ICS loss rate dominate firmly over the acceleration or is
negligible the energy distribution of outgoing electrons is
quasi-monoenergetic around the value which is well approximated with the
continuous approach.

We find that preventing electrons from achieving large Lorentz factors
occurs also for accelerating potential by Harding \& Muslimov (\cite{hm}) 
in its nonsaturated version with $\hacc\ll\rpc$.
The discrepancy between the Monte Carlo and the integration method
also occurs within the high-$B$ and low-$T$ regime
and is especially pronounced for $\rth \sim \hacc$.   

\begin{acknowledgements}
This work was supported by KBN grants 2P03D 01016 and 2P03D 02117.
JD thanks T.Bulik for useful advice in programming.
We appreciate  sugestions by the anonymous referee and  comments by 
Bing Zhang on the
manuscript which helped to improve the paper. 
\end{acknowledgements}

\newpage

\end{document}